\documentclass[prl,twocolumn,showpacs,amsmath,amssymb]{revtex4}
\usepackage[dvips]{graphicx}
\usepackage{bm}% bold math

\begin{document}
\title{Full Counting Statistics for a Single-Electron Transistor,\\
Non-equilibrium Effects at Intermediate Conductance}
\author{Yasuhiro~Utsumi$^{1,2}$, Dmitri~S.~Golubev$^{1,3,4}$, 
and Gerd~Sch\"on$^{1,4}$}
\affiliation{
$^1$Institut f\"{u}r Theoretische Festk\"{o}perphysik, Universit\"{a}t Karlsruhe, 76128 Karlsruhe, Germany \\
$^2$Condensed Matter Theory Laboratory, RIKEN, Wako, Saitama 351-0198, Japan \\
$^3$I.E. Tamm Department of Theoretical Physics, P.N. Lebedev Physics Institute, 119991 Moscow, Russia\\
$^4$ Forschungszentrum Karlsruhe, Institut f\"ur Nanotechnologie,
76021 Karlsruhe, Germany}

\begin{abstract}
We evaluate the current distribution for a single-electron transistor with intermediate strength tunnel conductance. Using the Schwinger-Keldysh approach and the drone (Majorana) fermion representation we account for the renormalization of system parameters. Nonequilibrium effects induce a lifetime broadening of the
charge-state levels, which suppress large current fluctuations.
\end{abstract}

\date{\today}
\maketitle

\newcommand{\rd}{d}
\newcommand{\ri}{i}
\newcommand{\mat}[1]{\mbox{\boldmath$#1$}}
\newcommand{\mtau}{\mbox{\boldmath$\tau$}}
\newcommand{\cgf}{{\cal W}}

%
%	full counting statistics
%
The `Full Counting Statistics' (FCS) of charge transport 
has proven to be a powerful tool in the description of current
fluctuations~\cite{Qua-Noi-Mes-Phy}.  
The concept had been explored by Levitov and Lesovik~\cite{Levitov1},
who expressed the FCS of an arbitrary mesoscopic structure with
non-interacting electrons in terms of its $S-$matrix.  
Much less is known about the FCS of interacting mesoscopic systems, 
a problem which has been addressed only
recently~\cite{KT,KG,bn,Bagrets1,Braggio}.   

As a fundamental example of an interacting mesoscopic systems we consider 
a Single-Electron Transistor (SET). It consists of a metallic island
coupled to drain and source (left and right) electrodes via
low-capacitance tunnel junctions, with  resistances $R_{\rm L}$ and
$R_{\rm R}$,  
as well as to a gate electrode. The  strength of Coulomb interaction
is characterized by the charging energy $E_C \! = \! e^2/2C_\Sigma$,
which depends on the total capacitance   
$C_\Sigma \! = \! C_{\rm L} \! + \! C_{\rm R} \! + \! C_{\rm G}$.  
A measure for the tunneling strength is the dimensionless parameter 
$\alpha_0 \! =\! 
(R_{\rm L} \! + \! R_{\rm R})/2\pi e^2 R_{\rm L}R_{\rm R}$
(we put $\hbar \!=\! k_{\rm B} \!=\! 1$). 

In Refs.~\cite{KT,KG} the FCS of a similar system - a quantum dot -
has been studied, fully accounting for strong electron correlations,
however only for a particular setup and parameters, corresponding to
the Toulouse point.  A renormalization group approach 
had been developed for the regime $\alpha_0 \! \gg \! 1$~\cite{bn}. 
In the opposite limit, $\alpha_0 \! \rightarrow \! 0$,  the FCS 
has been analyzed to lowest order in Ref.~\cite{Bagrets1} and
next-to-lowest order (cotunneling) in Ref.~\cite{Braggio}.
However, effects of quantum fluctuations induced by 
nonvanishing $\alpha_0$ are still unknown. 
The aim of the present paper is to derive the FCS for a SET beyond
perturbation theory in the intermediate strength tunneling regime
$\alpha_0 \! \lesssim \! 1$. 

Let us further specify the situation to be considered.
At low transport voltages and temperatures, $eV,T \! \ll \! E_C$, 
due to Coulomb blockade tunneling is suppressed in a SET, everywhere
except near specific values of the gate voltage, e.g., near 
$Q_{\rm G} \equiv C_{\rm G}V_{\rm G}=e/2$.
In the neighborhood of this conductance peak the Coulomb
barrier is $\Delta_0=E_C(1-2Q_{\rm G}/e)$.
For $\alpha_0 \ll 1$ electrons tunnel via the island sequentially 
only when $\mu_{\rm R} < \Delta_0 <\mu_{\rm L}$, where 
$\mu_{\rm L/R}  =  \kappa_{\rm L/R} eV$ 
is the voltage drop between the L/R electrode and the island,
and $\kappa_{\rm L/R}  =  \pm C_{\rm R/L}(C_{\rm L}  +  C_{\rm R})^{-1}$. 
With increasing $\alpha_0$, higher order effects such as cotunneling and
quantum fluctuations of the charge  gain importance~\cite{Golubev}.
They lead to a renormalization of $\Delta_0$ and $\alpha_0$. 
The perturbative renormalization group analysis~\cite{Matveev}
(for $eV\!=\!0$) predicts a renormalization factor $z_0  =  1/ \{ 1
+   2 \alpha_0 \ln(E_C/\Lambda) \}$ to depend logarithmically on the
cutoff energy $\Lambda=\max \{ \Delta_0,T \}$.  

{\it The model.} --
We concentrate on the tunneling regime with inverse $RC$ time 
$1/R_{\rm T} C_{\Sigma}
\!=\!
4 \pi \alpha_0 E_C$ 
smaller than $E_C$, which ensures that the charge-state levels are
well resolved.  
In the vicinity of the conductance peak, precisely for
$|\Delta_0|/E_C\ll 1$,  it is sufficient to restrict attention to
two charge states with charges differing by $e$.
The Hamiltonian can then be mapped onto  
the \lq multi-channel anisotropic Kondo model'~\cite{Matveev}.
Introducing a spin-1/2 operator $\hat{\sigma}$ acting on the
charge states,  we write 
%----------------------------------------------------------
\begin{eqnarray}
\hat{H}
 = &
\sum_{\rm r=L,R,I}
\sum_{k n}
\varepsilon_{{\rm r} k} \,
\hat{a}_{{\rm r} k n}^{\dag} \hat{a}_{{\rm r} k n}
+ 
\frac{\Delta_0}{2} \hat{\sigma}_z
\nonumber \\
& + 
\sum_{\rm r=L,R}
\sum_{k k' n}
(
T_{\rm r} 
\hat{a}_{{\rm I} k n}^{\dag} 
\hat{a}_{{\rm r} k' n}
\hat{\sigma}_{+}
+
{\rm H. c.}).
\label{eqn:H}
\end{eqnarray}
%----------------------------------------------------------
Here $\hat{a}_{{\rm r} k n}^{\dag}$ creates  
an electron with wave vector $k$ and channel index (including spin) $n$
in the left or right electrode or island (r=L,R,I). 
Tunneling matrix elements $T_{\rm r}$ are assumed to be
independent of $k$ and $n$. 
The junction conductances are 
$1/R_{\rm r} 
 =  2 \pi e^2 N_{\rm ch} |T_{\rm r}|^2 \rho_{\rm I} \rho_{\rm r}$,
with $N_{\rm ch}$ being the number of channels and $\rho_{\rm r}$
the electron DOS.
We implicitly assume that energy and spin relaxation times 
are fast, and electrons obey the Fermi distribution.

% drone-fermion representation & Schwinger-Keldysh & cumulants generating functional

A convenient tool to treat the spin-1/2 operators in Eq.~(\ref{eqn:H}) 
is the `drone' (Majorana) fermion representation~\cite{Isawa}, 
%----------------------------------------------------------
$\hat{\sigma}_{+}   =   \hat{c}^{\dag} \hat{\phi}$, 
$\hat{\sigma}_{z}   =   2 \hat{c}^{\dag} \hat{c} - 1$.
%----------------------------------------------------------
where 
$\hat{\phi}   =   \hat{d}^{\dag} + \hat{d}$
is a Majorana fermion and 
$\hat{c}$ and $\hat{d}$ are Dirac fermions. 
This formulation enables one to apply  Wick's theorem and 
the fermionic Schwinger-Keldysh approach~\cite{Chou,Kamenev}.

%
% Schwinger-Keldysh approach
%
{\it Cumulant generating functional.} --
The central object of our approach is the generating functional 
of connected Green's functions (GFs)
%----------------------------------------------------------
\begin{equation}
W[\varphi]   \equiv   -\ri \ln \int D 
[a_{{\rm r} k n}^*,  a_{{\rm r} k n},  c^*, c, d^*, d ]\,
e^{  \ri \int_C \rd t {\cal L}(t) }.
\label{eqn:generating_function}
\end{equation}
%----------------------------------------------------------
Here ${\cal L}$ is the Lagrangian corresponding to $\hat{H}$ 
(\ref{eqn:H}),
and six Grassmann variables have been introduced
(see Ref.~\cite{Utsumi} for details).
The closed time-path $C$ (Keldysh contour) runs from
$t = -\infty$ to $\infty$, back to $-\infty$, and 
connects to the imaginary time-path to end at $t = -\infty-\ri/T$. 
We introduce auxiliary source fields, 
the phase of the tunneling matrix element, 
$T_{\rm r} \rightarrow 
T_{\rm r} e^{i \kappa_{\rm r} \varphi(t)}$, 
distinguishing between
forward and backward time-paths, $\varphi_{+}(t)$ and $\varphi_{-}(t)$. 
The `center-of-mass' variable 
$\varphi_c(t)  \equiv  \{ \varphi_{+}(t)  +  \varphi_{-}(t) \}/2 = eVt$
is then fixed by the transport voltage.

From the Cumulant Generating Functional (CGF) 
one finds the number of transmitted electrons $q$ 
during the measurement time $t_0$, 
$\cgf(\lambda) =\sum_{n=0}^\infty
\langle \langle \delta q^n \rangle \rangle
(\ri \lambda)^n/n!$.
Following Ref.~\cite{Kamenev}, it is derived from
Eq.~(\ref{eqn:generating_function}) 
by fixing during the measurement the `counting field' (relative variable)
$\varphi_\Delta(t)  \equiv  \varphi_{+}(t)  -  \varphi_{-}(t)$ 
at a constant value $\lambda$: 
%----------------------------------------------------------
\begin{equation}
\cgf(\lambda) = 
\ri W[\varphi]|_{\varphi_c(t) = eVt,
\varphi_\Delta(t) = \lambda \theta(t_0/2+t) \theta(t_0/2-t)}.
\label{eqn:Kamenevs}
\end{equation}
%----------------------------------------------------------
The distribution of $q$ (or equivalently of the current $I  \equiv e q/t_0$) 
is given by the inverse Fourier transformation, 
%----------------------------------------------------------
\begin{equation}
P = 
\frac{1}{2 \pi}
\int^{\pi}_{-\pi}       
\rd \lambda 
\,
e^{\cgf(\lambda) -\ri q \lambda}
\approx
e^{\cgf(\lambda^*) - \ri (t_0 I/e)  \lambda^*}. 
\label{eqn:invFourier}
\end{equation}
%----------------------------------------------------------
The integral can be evaluated in saddle point approximation, with
$\lambda^*$ following from the relation  
$I  =  -\ri e \partial_\lambda \cgf(\lambda^*)/t_0$~\cite{Bagrets1}. 
This approach is valid for long measurement times 
since $\cgf$ is proportional to $t_0$ (see below).

We proceed following Ref.~\cite{Utsumi},
where a conserving approximation
for the second cumulant had been developed.
Tracing out the electron degrees of freedom leads to
an effective action for the $c$ and $d$ fields,
$S^\lambda   \equiv  S_{\rm ch}   +   S_t^\lambda$, composed of a charging 
and a tunneling term:
%----------------------------------------------------------
\begin{eqnarray}
 S_{\rm ch}
 =  
\int _C   \rd t \{ c(t)^* (\ri \partial_t -  \Delta_0) c(t) + 
\ri d(t)^* \partial_t d(t) \}, \hspace{0.6cm}
\\
 S_t^\lambda
=
-   
\int _C     \rd t \rd t' \,
c^*(t) \phi(t)
\, \alpha^\lambda(t,t') \,
\phi(t') c(t')
+O(T_{\rm r}^4). 
\label{eqn:tunnelaction}
\end{eqnarray}
%---------------------------------------------------------- 
Here
$\alpha^\lambda  =  \alpha_{\rm L}^\lambda 
 +  \alpha_{\rm R}^\lambda$ 
is a particle-hole GF describing 
tunneling of an electron from one electrode to the island. 
It depends on the counting field. 
The connection to the ordinary GF 
is established by a rotation 
by $\lambda_{\rm r}  =  \kappa_{\rm r} \lambda$
in the Keldysh space as follows:
%----------------------------------------------------------
\begin{equation}
\tilde{\alpha}^{\lambda}_{\rm r}(\omega)
 = 
U_{\lambda_{\rm r}}^\dagger 
\tilde{\alpha}_{\rm r}(\omega)
U_{\lambda_{\rm r}}, 
\,
\tilde{\alpha}_{\rm r}(\omega)
 = 
\biggl(  
\begin{array}{cc}
0 &   \alpha_{\rm r}^A(\omega) \\
\alpha_{\rm r}^R(\omega) &   \alpha_{\rm r}^K(\omega)
\end{array}
\biggl),
\label{eqn:p-h}
\end{equation}
%----------------------------------------------------------
where 
$U_{\lambda_{\rm r}}
 = 
\exp(- \ri \lambda_{\rm r} \mtau_1/2)$
and
$[\mtau_1]_{ij}  =  1   -   \delta_{ij}$.
The retarded and advanced components are given by 
$$
\alpha^R_{\rm r}(\omega) 
 =  
\alpha^A_{\rm r}(\omega)^*
 = 
-\ri \pi 
\alpha_0^{\rm r}\, \frac{ 
(\omega  -  \mu_{\rm r})E_C^2}{
(\omega  -  \mu_{\rm r})^2 + E_C^2 },
$$
where $\alpha_0^{\rm r}=1/2 \pi e^2 R_{\rm r}$, and the Keldysh component by
$\alpha^K_{\rm r}(\omega)  =  
2 \, \alpha^R_{\rm r}(\omega) \coth(\omega-\mu_{\rm r})/2T.$
We introduced a Lorentzian cutoff 
to regularize the ultraviolet divergence
and ignored the term $O(T_{\rm r}^4)$ in the action (\ref{eqn:tunnelaction}),  
since it is small in the limit 
$N_{\rm ch} \! \gg \! 1$. 

The free retarded GF of the Dirac fermion $\hat{c}$, 
$g_c^R(\omega)  =  1/(\omega + \ri \eta - \Delta_0)$, 
describes the dynamics of charge  excitations 
($\eta$ is a positive infinitesimal). 
The corresponding self-energy 
$\Sigma_c^\lambda  =  
\Sigma_{\rm L}^\lambda  +  \Sigma_{\rm R}^\lambda$ 
accounts for quantum fluctuations of the island charge caused by tunneling. 
Integrating out $d$-fields, we obtain the components
of the self-energy in first order in $\alpha_0$:
%----------------------------------------------------------
\begin{equation}
\Sigma^K_{\rm r}   (\omega)  =  2 \alpha^R_{\rm r}(\omega),
\;\;
\Sigma^R_{\rm r}(\omega)
 =  \int    \frac{\rd \omega'}{2\pi}\, 
\frac{\ri \alpha^K_{\rm r}(\omega')}
{ \omega + \ri \eta - \omega' }. 
\end{equation}
%----------------------------------------------------------
(For simplicity we present here only the result for $\lambda\!=\!0$.)
For a symmetric SET ($R_{\rm L} \! = \! R_{\rm R}$,  
$C_{\rm L} \! = \! C_{\rm R}$), at $T\!=\!0$
and  $|\omega| \! \ll \! eV$, one finds $\Sigma_c^R(\omega)  \approx  
\alpha_0 \ln ( 2 E_C/eV ) \, \omega - \ri \Gamma/2$,
where
%----------------------------------------------------------
$\Gamma   =  
\Gamma_{  \rm I L}  +  \Gamma_{  \rm L I} 
 + 
\Gamma_{  \rm I R}  +  \Gamma_{  \rm R I}$
%----------------------------------------------------------
is the sum of the rates 
%----------------------------------------------------------
$\Gamma_{  \rm rI/Ir}  = \pm(1/e^2   R_{\rm r}) \,
(\Delta_0   -   \mu_{\rm r})/(e^{\pm(\Delta_0   -   \mu_{\rm r})/T} - 1) $ 
%----------------------------------------------------------
describing tunneling into (out of) the island 
through the junction r, evaluated by Fermi's golden rule.

We can proceed in a systematic diagrammatic expansion 
in $\alpha_0$~\cite{Utsumi}.
To lowest order one obtains for the CGF:
%----------------------------------------------------------
$\cgf^{[1]}(\lambda)  =  -   \int_C   \rd t \rd t'
g_c(t,t') \Sigma^\lambda_c(t',t)$. 
%----------------------------------------------------------
We project the time from the Keldysh contour $C$ to the real
axis and observe that  for long enough measurement times 
we can approximate 
%----------------------------------------------------------
$\delta_{t_0} (\omega)
\! \equiv \! 
\int^{t_0/2}_{-t_0/2} \rd t \, e^{-\ri \omega t}/2 \pi$
%----------------------------------------------------------
by a $\delta$-function, 
$\delta_{t_0}(\omega) \to \delta(\omega)$, and
$ \big(\delta_{t_0} (\omega)\big)^2 \to t_0\delta(\omega)/2\pi$.
The latter ensures that any closed diagram, 
consequently $\cgf$, is proportional to $t_0$. 
After Fourier transformation we obtain
%----------------------------------------------------------
\begin{eqnarray}
&& \cgf^{[1]}(\lambda)
  \approx  
- t_0   \int   \rd \omega {\rm Tr}
\{ \tilde{g}_c(\omega) \mtau_{  1} 
\tilde{\Sigma}^\lambda_c(\omega) \mtau_{  1} \} /2 \pi
\nonumber \\
&&  =  
t_0 
\sum_{\rm r=L,R}
\{ 
P_- \Gamma_{  \rm r I} (e^{\ri \lambda_{\rm r}} - 1)
 + 
P_+ \Gamma_{  \rm I r} (e^{-\ri \lambda_{\rm r}} - 1) \}. 
\label{eqn:first}
\end{eqnarray}
%----------------------------------------------------------
Here we used the expression 
for the Keldysh component of $c$-field GF, 
$g_c^K(\omega) \! = \! 2 \ri \, {\rm Im} \, 
g_c^R(\omega)(P_{  -} \!  - \!  P_{  +})$, which 
contains equilibrium occupation probabilities of the charge states
$Q_{\rm G}$ and $Q_{\rm G}-e$:
$P_{  \pm}  =  1/(e^{\pm \Delta_0/T} \! + \! 1)$.

At this point we note that the naive 
first order expansion~(\ref{eqn:first}) is insufficient.
First, it contains the equilibrium occupation probabilities
rather than the stationary ones. 
Second, due to charge conservation the CGF should depend only on
the difference of the counting fields 
$\lambda_{\rm L}  -  \lambda_{\rm R}  =  \lambda$~\cite{Belzig},
which is also violated. 
These problems are resolved if we sum up an infinite subclass of diagrams. 
Specifically, we sum up the geometric series in 
$(\tilde{g}_c \mtau_{  1} \tilde{\Sigma}^\lambda_c \mtau_{  1})$,
which contains the leading logarithms, 
i.e. powers of $\alpha_0 \ln( 2 E_C/eV)$, and get 
%----------------------------------------------------------
\begin{eqnarray}
&& \cgf(\lambda)
=
t_0
   \int    \frac{\rd \omega}{2\pi}
\, {\rm Tr} \, 
\ln   \left[
{\tilde{g}_c(\omega)}^{-1}
   -
\mtau_{  1}
\tilde{\Sigma}_c^\lambda(\omega)
\mtau_{  1}
\right]
\nonumber
\\ &&
=\,
t_0
   \int   \frac{ \rd \omega}{2\pi}
\ln
\left[
1+
T^F (\omega) 
f (\omega  -  \mu_{\rm L}) 
h (\omega  -  \mu_{\rm R}) 
(e^{\ri \lambda} - 1)
\right.
\nonumber \\ &&
\left. 
+\,
T^F (\omega) 
f (\omega  -  \mu_{\rm R}) 
h (\omega  -  \mu_{\rm L})
(e^{-\ri \lambda} - 1)
\right].
\label{eqn:CGF-RTA}
\end{eqnarray}
%----------------------------------------------------------
Here 
$f(\omega) \!=\! 1/(e^{\omega/T} \!+\! 1)$,  
$h(\omega)  =  1 - f(\omega)$ and
\begin{equation}
T^F   (\omega)
  =   -\alpha_{\rm L}^K   (\omega) \alpha_{\rm R}^K   (\omega)/
|\omega - \Delta_0 - \Sigma_c^R(\omega)|^2.
\label{TF}
\end{equation}
Note that we subtracted a constant from the CGF in order to satisfy
the normalization condition $\cgf(0) = 0$. 

Eq.~(\ref{eqn:CGF-RTA}) is the main result of this paper. 
It is similar to  
Levitov-Lesovik formula~\cite{Levitov1}, 
but the effective transmission probability (\ref{TF}) 
accounts for quantum fluctuations of the charge.
Using the condition Eq.~(\ref{eqn:Kamenevs}), Eq.~(\ref{eqn:CGF-RTA})
can also be obtained from an approximate $W$ given in 
Eq.~(25) of Ref.~\cite{Utsumi}. 
Thus, the first and second cumulants, 
%----------------------------------------------------------
$\langle I \rangle 
 = 
e \langle \langle \delta q \rangle \rangle/t_0$ 
%----------------------------------------------------------
and 
%----------------------------------------------------------
$S_{ I I}  =  
2 e^2 \langle \langle \delta q^2 \rangle \rangle/t_0$,
%----------------------------------------------------------
reproduce the average current~\cite{SchoellerSchoen} and
zero-frequency noise~\cite{Utsumi}, derived before. 

Although we used only the first order expansion for the self-energy,
Eq.~(\ref{eqn:CGF-RTA}) is exact to second order in $\alpha_0$. 
One can check that the diagrams
ignored in Eq.~(\ref{eqn:CGF-RTA}) within
second order expansion, 
i.e. the diagrams with intersecting interaction lines, 
are proportional to $\Delta_0/E_C\ll 1$.
Furthermore, higher order terms of Eq.~(\ref{eqn:CGF-RTA}) generate the renormalization 
factor $z_0$ 
consistent with the renormalization group result~\cite{Matveev}.  

{\it Limiting cases.} --
In the limit $\alpha_0 \! \rightarrow \! 0$,  
Eq.~(\ref{eqn:CGF-RTA}) reproduces
the result of the `orthodox' theory~\cite{Bagrets1}:
%----------------------------------------------------------
\begin{eqnarray}
\cgf^{(1)}(\lambda)
  =  
t_0 \Gamma  \frac{\sqrt{D(\lambda)}-1}{2},\hspace{3.4cm}
\\
D(\lambda)
 =  
1 +
\frac{4\Gamma_{\rm L I} \Gamma_{\rm I R}(e^{i \lambda} \!-\! 1)}{\Gamma^2}
+
\frac{4\Gamma_{\rm R I} \Gamma_{\rm I L}(e^{-i \lambda} \!-\! 1)}{\Gamma^2}.
\end{eqnarray}
%----------------------------------------------------------
%
% second order perturbation theory
%
The second order expansion in $\alpha_0$ reads, 
%----------------------------------------------------------
\begin{equation}
\cgf^{(2)}(\lambda)   =   \partial_{\Delta_0} \{{\rm Re} \Sigma^R_c(\Delta_0) \cgf^{(1)}(\lambda) \}
+ \cgf^{\rm cot}(\lambda). 
\label{eqn:CGF-2nd}
\end{equation}
%----------------------------------------------------------
The first term of this equation provides the renormalization 
of the system parameters up to first order in $\alpha_0$.
Namely, the parameters are renormalized as
$\alpha_0 \! \to \! 
\alpha_0 \{1 \! + \! \partial_{\Delta_0} {\rm Re} \Sigma^R_c(\Delta_0) \}$
and 
$\Delta_0 \! \to \! \Delta_0 \! + \! {\rm Re} \Sigma^R_c(\Delta_0)$. 
This agrees with the corresponding results obtained
earlier for the average current~\cite{KSS}. 
It is also consistent with the recent results of Braggio {\sl et
al.}~\cite{Braggio}. We also checked that Eq.~(\ref{eqn:CGF-2nd}) 
can be reproduced by the systematic real-time diagrammatic 
expansion similar to that of Ref.~\cite{SchoellerSchoen}.

The second term of Eq.~(\ref{eqn:CGF-2nd}) is the CGF of a
bidirectional Poissonian process 
%----------------------------------------------------------
\begin{equation}
\cgf^{\rm cot}(\lambda)   =  
t_0 \{ 
\gamma^+ (e^{i \lambda} - 1) 
+ 
\gamma^- (e^{-i \lambda} - 1) \}, 
\end{equation}
%----------------------------------------------------------
governed by the cotunneling rates 
%----------------------------------------------------------
\begin{eqnarray} 
\gamma^\pm = 
\int  \rd \omega
\frac{2 \pi \,
\alpha_0^{\rm L} \, \alpha_0^{\rm R}\;(\omega  -  \mu_{\rm L}) 
(\omega  -  \mu_{\rm R})}
{(e^{\pm \frac{\omega-\mu_{\rm L}}{T}}- 1)(1-e^{\mp \frac{\omega-\mu_{\rm R}}{T}})}
%{(e^{\pm (\omega-\mu_{\rm L})/T}- 1)(e^{\mp (\omega-\mu_{\rm R})/T}- 1)} 
%\nonumber\\ &&\times\,
{\rm Re}  \frac{1}{(\omega + \ri \eta - \Delta_0)^2}.
\nonumber
\end{eqnarray}
%----------------------------------------------------------
This term is relevant in the Coulomb blockade regime and is consistent with 
the FCS theory of quasiparticle tunneling~\cite{Levitov2}. 

%
% Plot
%

%===========================================================
\begin{figure}[ht]
\includegraphics[width=1.0 \columnwidth]{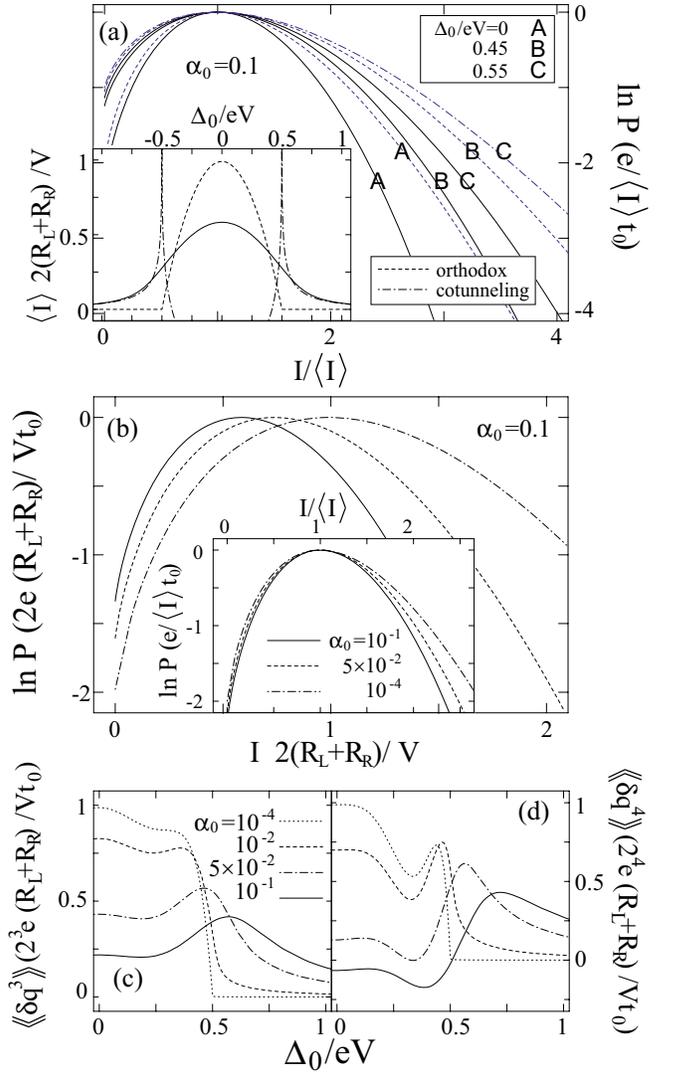}
\caption{
Current distribution $P$ of a symmetric SET 
[$T=0$ and $eV/E_C=0.2$].
(a) Solid lines are plots of $P$ for various values of $\Delta_0$. 
The dashed lines represent
the `orthodox' theory and the dot-dashed line 
represents the cotunneling expansion. 
The inset shows the average current for the same parameters.\\
(b) Plot of $P$ at $\Delta_0  = 0$ for various values
of the conductance  
as a function of the current normalized to $V/2 (R_{\rm L}+R_{\rm R})$;
inset: the same distribution normalized to 
the average current $\langle I \rangle$.\\
(c) The skewness and (d) kurtosis 
for various conductances. 
}
\label{fig:distribution}
\end{figure}
%===========================================================

% 
% orthodox - cotunneling
% 

At the conductance peak, $\Delta_0  = 0$, for a symmetric SET  and $T=0$, 
the `orthodox' theory yields  
%----------------------------------------------------------
$\cgf^{(1)} \approx 
2 \, \bar{q} \, (e^{\ri \lambda/2} - 1)$, 
%----------------------------------------------------------
where $e \bar{q}/t_0  =  V/2(R_{\rm L}   +   R_{\rm R})$.
The factor $e^{\ri \lambda/2}$ leads to 
a sub-Poissonian value of the Fano factor 
$S_{II}/2 e \langle I \rangle   \approx   1/2$,
indicating that tunneling processes are correlated. 
The origin of this correlation can be understood
from the explicit form of the distribution 
%----------------------------------------------------------
$P(q)  =  
\sum_{q_{\rm L},q_{\rm R}=0}^\infty
P_{\rm P}(q_{\rm L}) 
P_{\rm P}(q_{\rm R}) 
\, \delta_{q,(q_{\rm L}+q_{\rm R})/2},$
%----------------------------------------------------------
obtained by inverse Fourier transformation 
of Eq.~(\ref{eqn:invFourier}) without saddle point approximation.
The numbers of electrons  transmitted 
through either junction, 
$q_{\rm L}$ and $q_{\rm R}$, follow 
the same Poissonian distribution 
%----------------------------------------------------------
$P_{\rm P}(q) 
 = 
\bar{q}^{\, q} e^{-\bar{q}}/q! \,$. 
%----------------------------------------------------------
The Kronecker delta implies that
$q_{\rm L}$ and $q_{\rm R}$ are correlated.

With $\Delta_0$ approaching the threshold value $\Delta_0/eV=0.5$, 
the tunneling onto the island becomes the bottleneck and 
the CGF acquires the Poissonian form, 
%----------------------------------------------------------
$\cgf^{(1)}
  \approx  
t_0 \Gamma_{  \rm L I} \, (e^{\ri \lambda} - 1)$. 
%----------------------------------------------------------
It remains Poissonian in the cotunneling regime, 
$|\Delta_0/eV| > 0.5$, 
%----------------------------------------------------------
$\cgf^{\rm cot}
  \approx  
t_0 \gamma^+ \, (e^{\ri \lambda} - 1)$. 
%----------------------------------------------------------

Let us compare our result~(\ref{eqn:CGF-RTA}) to  the `orthodox' and 
the cotunneling theories. 
The latter theories fail around the threshold $\Delta_0/eV = 0.5$. For example,
in the average current, shown in
the inset of Fig.~\ref{fig:distribution}(a), we observe 
a mismatch between their predictions  since  
$\Gamma_{  \rm L I} \rightarrow 0$ while 
$\gamma^+ \rightarrow \infty$.
In contrast, the distribution 
derived from Eq.~(\ref{eqn:CGF-RTA}), shown in 
Fig.~\ref{fig:distribution}(a) by solid lines, behaves regular. 
It widens with increasing $\Delta_0$. 
The `orthodox'  (dashed lines) and 
the cotunneling  (dot-dashed line) theories 
show the same trend but overestimate the width.

{\it Renormalization and lifetime broadening effects.} --
For large conductance, quantum fluctuations of the
charge are pronounced. 
However, as long as 
$z_0 \Gamma   \ll   \Lambda$, where 
$\Lambda = \max(|z_0 \Delta_0|,2 \pi T, |eV|/2)$, 
the `orthodox' CGF $\cgf^{(1)}$ with 
renormalized parameters $z_0 \alpha_0$ and $z_0 \Delta_0$
remains a good approximation.  This scenario fails in the regime
$\Lambda \ll T_K = E_Ce^{-1/2 \alpha_0}/ 2 \pi$
where the approximation of leading logarithms becomes insufficient.

The renormalization effect is illustrated in
Fig.~\ref{fig:distribution}(b), where
the current distribution at $\Delta_0  = 0$ is plotted.
Since $z_0$ decreases with increasing $\alpha_0$, 
the mean value of the current, i.e. a peak position, shifts to lower values. 
The renormalization effect can be absorbed when
we plot $\ln P$ with horizontal axis normalized
by $\langle I \rangle$ rather than $V/2(R_{\rm L}+R_{\rm R})$.
However, even after plotting the distribution as a function of 
the normalized current [inset of Fig.~\ref{fig:distribution}(b)]
the three curves do not collapse to a single one.
The remaining differences can be attributed 
to the non-Markovian lifetime broadening effect 
as described by ${\rm Im} \, \Sigma^R_c$. 
We observe that the current distribution shrinks with increasing $\alpha_0$. 
This agrees with the previously noted suppression of 
the Fano factor~\cite{Utsumi}. FCS provides further
information, showing in detail how 
the probability for currents exceeding the average value is suppressed.

The effect of lifetime broadening is also visible in the moments.
At moderately large voltages, $eV   \gg   T_{\rm K}$, and at $T=0$ 
the real part of the self-energy $\Sigma^R_c$ is negligible and 
$\Sigma^R_c(\omega)  \approx - \ri \pi \alpha_0 eV$. 
The CGF at $\Delta_0 = 0$ then is
%----------------------------------------------------------
\begin{eqnarray}
\cgf(\lambda) &\approx&
2 \, \bar{q}
\{
(e^{\ri \lambda/2} - 1)
-2 \alpha_0 (e^{\ri \lambda} - 1)
\nonumber \\
&+& 
\pi^2 \alpha_0^2 
(e^{\ri 3 \lambda/2} - e^{\ri \lambda/2})/2
+O(\alpha_0^3)
\}, 
\label{eqn:CGFexpansion}
\end{eqnarray}
%---------------------------------------------------------- 
and the ratio of higher order cumulants to the first one becomes
%----------------------------------------------------------
$
\langle \langle \delta q^n \rangle \rangle/
\langle \langle \delta q \rangle \rangle
 =  
2^{1-n} 
\{1  -  4 \alpha_0(2^{n-1}  -  1) 
+ O(\alpha_0^2) \}$.
%----------------------------------------------------------
We note that, as $\alpha_0$ increases, higher order cumulants are suppressed 
as compared to the Poissonian result $2^{1-n}$.

Figures~\ref{fig:distribution}(c) and (d) show
the skewness 
$\langle \langle \delta q^3 \rangle \rangle$
and the kurtosis 
$\langle \langle \delta q^4 \rangle \rangle$
as a function of $\Delta_0$. 
A peak  around the threshold develops with increasing conductance. 
We expect that this kind of behavior can be observed
with present-day experimental techniques~\cite{Reulet}. 

In conclusion, we have derived the Full Counting Statistics for a
single-electron transistor in the vicinity of a conductance peak.  
Quantum fluctuations of the charge are taken into account 
by a summation of a certain subclass of diagrams,
which corresponds to the leading logarithmic approximation. 
In first order in $\alpha_0$ our results reproduce 
the `orthodox' theory, while in second
order they account for renormalization and cotunneling effects. 
We have shown that in non-equilibrium situations quantum
fluctuations of the charge induce lifetime broadening for the
charge states of the central island.  
An important consequence is the suppression of the probability 
for currents larger than the average value.

% acknowledge

We thank 
D. Bagrets,
A. Braggio,
Y. Gefen,
J. K{\"o}nig,
A. Shnirman
for valuable discussions.
YU was supported by the 
DFG 
\lq\lq Center for Functional Nanostructures" 
and 
RIKEN Special Postdoctoral Research Program.

\end{document}